\documentclass[amsmath, amsonfts,twocolumn]{revtex4}

\def\minus{%
  \setbox0=\hbox{-}%
  \vcenter{%
    \hrule width\wd0 height \the\fontdimen8\textfont3
  }%
}

\usepackage{graphicx}
\usepackage{amssymb}
\usepackage{wasysym}
\usepackage{stmaryrd}
 
\usepackage{lineno,hyperref}
\usepackage{amsmath,bm}

\newcommand{\bvec}{\left(\begin{array}{c}}
\newcommand{\evec}{\end{array}\right)}
\newcommand{\bmat}{\left(\begin{array}{cc}}

\newcommand{\toutin}{\left\{\begin{array}{l}}
\newcommand{\toutind}{\left\{\begin{array}{ll}}
\newcommand{\toutout}{\end{array}\right.}

\renewcommand{\phi}{\varphi}

\def\beq{\begin{equation}}
\def\eeq{\end{equation}}

\begin{document}

\title{Low frequency acoustic stop bands in cubic arrays of thick spherical shells with holes}
\author{Guillaume Dupont}
\address{Aix Marseille Univ, CNRS, Centrale Marseille, IRPHE UMR 7342, 13013 Marseille, France
}
\author{Alexander Movchan}
\address{Department of Mathematical Sciences, Liverpool University, Peach Street, Liverpool L69 3BX, UK
}
\author{Stefan Enoch}
\address{Aix Marseille Univ, CNRS, Centrale Marseille, Institut Fresnel UMR 7249, 13013 Marseille, France
}
\author{S\'ebastien Guenneau}
\address{Aix Marseille Univ, CNRS, Centrale Marseille, Institut Fresnel UMR 7249, 13013 Marseille, France
}

\begin{abstract} 
We analyse the propagation of pressure waves within a fluid filled with a three-dimensional array
of rigid coated spheres (shells). We first draw band diagrams for corresponding Floquet-Bloch waves. We then dig a channel terminated by a cavity within each rigid shell and observe the appearance of a low frequency stop band. The underlying mechanism is that each holey shell now acts as a Helmholtz resonator supporting a low frequency localized mode: Upon resonance, pressure waves propagate with fast oscillations in the thin water channel drilled in each shell and are localized in each fluid filled inner cavity. The array of fluid filled shells is approximated by a simple mechanical model of springs and masses allowing for asymptotic estimates of the low frequency stop band. We finally propose a realistic design of periodic macrocell with a large defect surrounded by 26 resonators connected by thin straight
rigid wires, which supports a localized mode in the low frequency stop band.
\end{abstract}

\maketitle

\section{\label{sec:level1}Introduction: Acoustic metamaterials}

In the tracks of photonic crystals, phononic crystals \cite{dowling} have provided a fillip
for research in acoustic stop band structures \cite{prb1999,mmp2002} within which light or sound
is prohibited to propagate due to multiple scattering between periodically spaced inclusions.
In 2000, Liu {\it et al.} provided the first numerical and
experimental evidence of locally resonant structures for elastic
waves in 3D arrays of thin coated spheres \cite{pingshen} wherein low frequency stop
bands occur. This seminal work paved the way towards acoustic analogues of electromagnetic meta-materials,
such as fluid-solid composites \cite{mei06}. Inspired by the research monograph on multi-structures
\cite{kmm1999}, Movchan and Guenneau subsequently proposed to use arrays of
cylinders with a split ring cross section as building blocks for 2D
localised resonant acoustic structures displaying negative
refraction \cite{mg,gmrp2007}. Such split ring resonators, introduced by John Pendry
in the context of electromagnetic waves \cite{pendry}, also work for in-plane elastic waves
\cite{gmm2007}. Milton, Briane and
Willis provided a thorough mathematical frame for such effects
including cloaking for certain types of elastodynamic waves in
structural mechanics \cite{graeme}. For instance, coupled in-plane
pressure and shear waves were numerically shown to be detoured
around a finite size obstacle by a specially designed cloak with
an anisotropic heterogeneous elasticity tensor (without
the minor symmetries) \cite{michele}. 
Acoustic metamaterials via geometric transform
can thus in theory achieve unprecented control of elastic and
pressure waves \cite{norris1,norris2}.

Li and Chan independently proposed a similar type of negative acoustic metamaterial \cite{li}. In a
recent work, Fang {\it et al.} experimentally demonstrated a dynamic effective negative stiffness
in a chain of water filled Helmholtz's resonators for ultrasonic waves \cite{fang}.
It has been
also shown using homogenisation theory that surface water waves propagating within
an array of fluid filled Helmholtz's resonators display a negative effective density
\cite{pre2009}. A focussing effect through a finite array of such resonators
was numerically achieved, with a resolution of a third of the wavelength.
Similar effects have been experimentally demonstrated for
a doubly periodic array of Helmholtz's resonators shaped as soda cans \cite{lerosey}.
In the present paper, we would like to extend these concepts to pressure
waves propagating in a three-dimensional array of resonators.

\section{\label{sec:level1} Motivation: Spectral properties of a periodic array of rigid spheres}

\begin{figure}[htp]
  \centering
\includegraphics[scale=0.5]{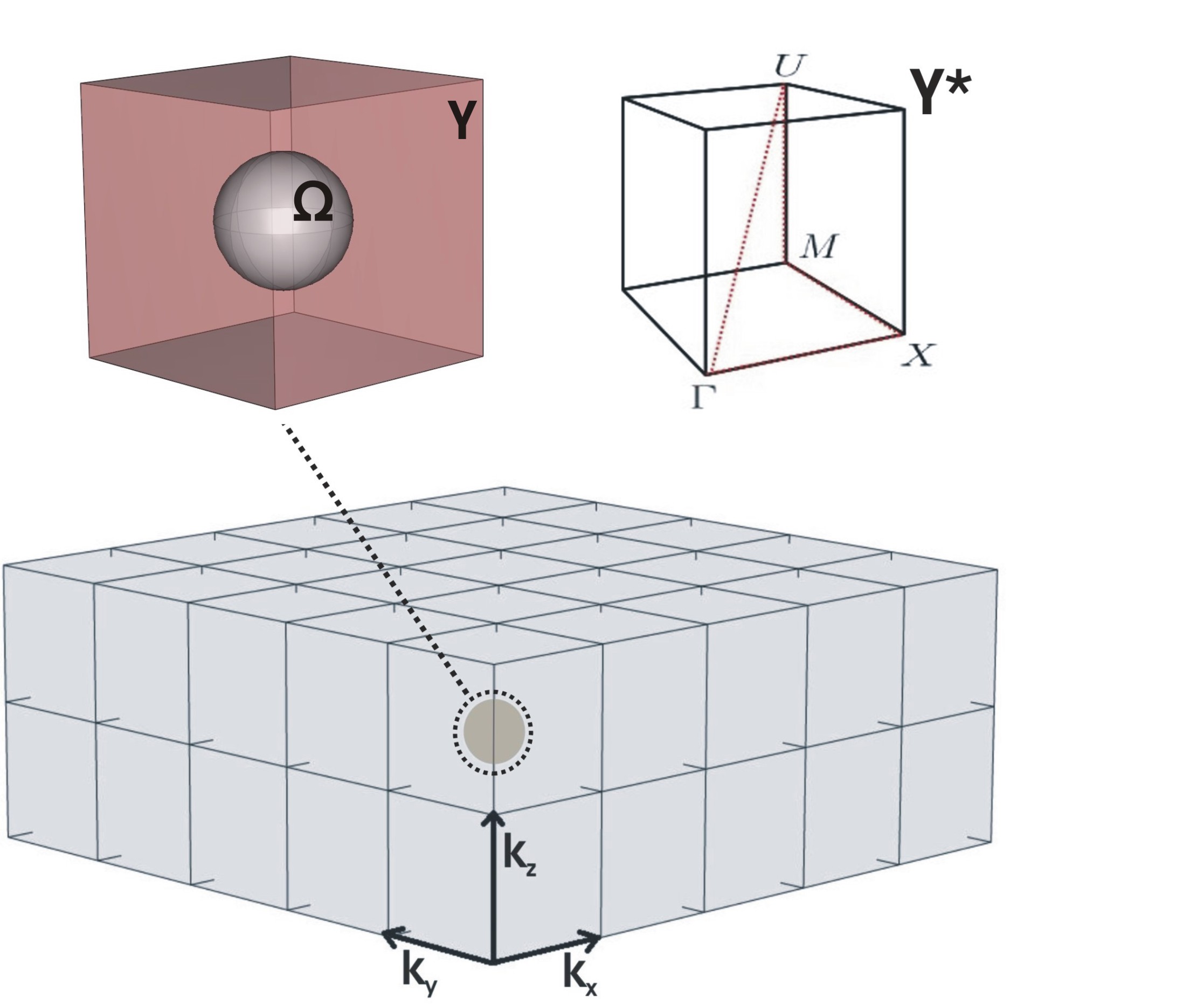}
\caption{Three-dimensionnal phononic crystal, periodic cell $Y$ with a rigid sphere $\Omega$
in physical space and
and irreducible Brillouin zone $\Gamma X M U$ of the periodic cell $Y^{*}$ in reciprocal space
with the three components of the Floquet-Bloch vector ${\bf k}=(k_x,k_y,k_z)$.}
\label{fig1}
\end{figure}

\begin{figure}[htp]
  \centering
\includegraphics[scale=0.17]{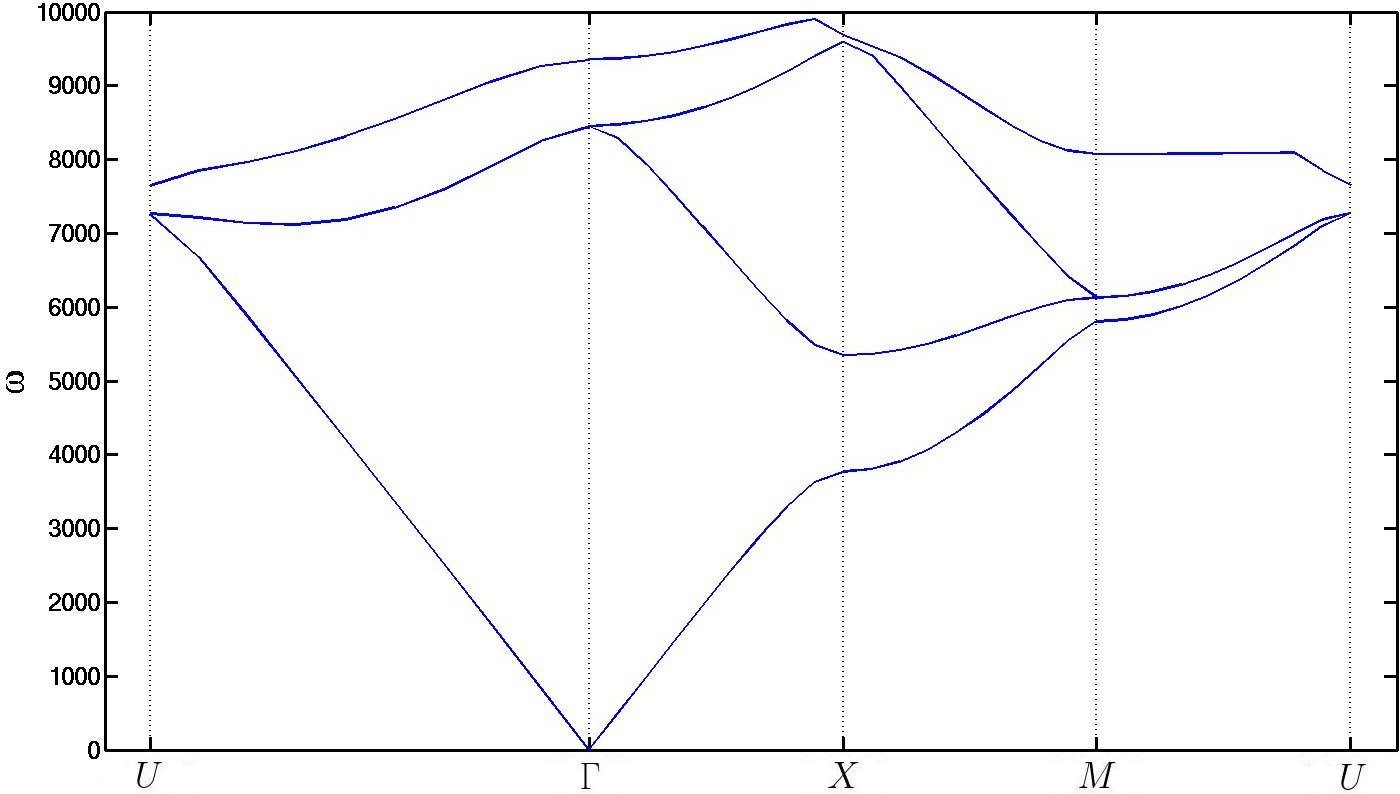}
\caption{Dispersion diagram for a periodic array (pitch $d=1$m) of spherical rigid inclusions ($R=0.4$m) representing the frequency 
$\omega$ (Hz)
of pressure waves in water, versus the wavenumber $|$\textbf{k}$|$ ($m^{-1}$), projection
of the Bloch vector ${\bf k}$ along the edges of $\Gamma X M U$.}
\label{fig2}
\end{figure}

As a preamble, let us start with an illustrative numerical result for a spectral problem for the Helmholtz operator within a periodic 
cubic array of rigid spheres: the unknown is a pressure wave field, here sound in water (wave speed $c=1483 \; m.s^{-1}$).
Neumann boundary conditions are prescribed on the contour of each defect and standard 
Floquet-Bloch conditions are set on an elementary cell of the periodic structure.
The finite element formulation was implemented in the
COMSOL Multiphysics Package to compute the eigenvalues and to 
generate the corresponding eigenfields. We present in figure \ref{fig1} the structure and in
figure \ref{fig2} the corresponding dispersion diagram for 
eigenfrequencies $\omega$ as a function of the Floquet-Bloch parameter $k$: along the horizontal axis we have 
the values of modulus of \textbf{k}, where $k$ stands for the position vector of a point on the contour 
$\Gamma X M U$ within the irreducible Brillouin zone.
We note the absence of bandgaps with the presence 
of rigid spheres. This lack of intervals of forbidden frequencies  motivates the present study: how can one create
a stop band without further increasing the size of the rigid spheres?

\section{\label{sec:level1}Setup of the spectral problem: The continuum model}
Let us first recall  the finite element set-up. Let
$u(x,y,z)$ satisfy the Helmholtz equation:

\begin{equation}
 \nabla . \left( \dfrac{1}{\rho(x,y,z)} \nabla u(x,y,z) \right) + \dfrac{\omega^2}{\lambda(x,y,z)}u(x,y,z) = 0 \label{eq1}
\end{equation}
where $\rho$, $\lambda$ and $u$ are the density, the bulk modulus and 
the pressure field.\\

\noindent Due to the periodicity of the lattice, it is customary to require that the eigenfunctions be of the 
Floquet-Bloch type. So, for a cubic array of unit cells $Y$,

\begin{equation}
u(x+1,y+1,z+1) = u(x,y,z)e^{i(k_x+k_y+k_z)}
\end{equation}
where $k_x$, $k_y$ and $k_z$ are components of the Bloch vector \textbf{k} within the Brillouin zone 
$Y^* = [0,\pi]^3$.\\

\noindent The implementation in the finite element package is fairly straightforward. We first multiply equation (\ref{eq1}) 
by a smooth function $V$ and using the Green's formula, we obtain the so-called weak form
of the time-harmonic acoutic equation


\begin{multline}
 -\int_Y \rho^{-1} \nabla u . \nabla V dx dy dz + \int_{\partial Y} \rho^{-1} \left( \dfrac{\partial u}{\partial n}V
-\dfrac{\partial V}{\partial n}u \right) ds \\
+\omega^2 \int_Y \lambda^{-1}uV dx dy dz = 0 \label{eq3}
\end{multline}
where $\partial f/\partial n=\nabla f\cdot {\bf n}$ with ${\bf n}$ the unit outward normal to the boundary
$\partial Y$ of $Y$, and $ds$ the infinitesimal surface element on $\partial Y$.

\noindent We note that the weak formulation holds for heterogeneous fluids as $\rho$ and $\lambda$
can be spatially varying. In particular, this model works for perforated domains such as a
homogeneous fluid filled with a periodic array of rigid osbtacles. For the finite
element implementation, (\ref{eq3}) is discretised using test functions
taking values on nodes of a tetrahedral mesh of the basic cell  (first order tetrahedral elements), see e.g.
\cite{andre} for further details. From (\ref{eq3}), we note that setting rigid boundary conditions conditions on an inclusion amounts
to assuming Neumann (natural) homogeneous data, whereas transmission conditions at the interface between
various fluid phases mean that the quantity $\rho^{-1}\partial u/\partial n$ is preserved across the interface.


\noindent Let us now consider a periodic array of defects $\Omega_1,...,\Omega_N$ embedded in an elementary cell $Y = ]0;1[^3$. Let $u(x,y,z)$ 
satisfy the Helmholtz equation in $Y\setminus \bigcup_{j=1}^N\overline{\Omega_j}$.
We also assume that $u$ satisfies Neumann boundary condiditons on the contours of defects, where ${\bf n}$ denotes the 
unit outward normal to the boundary $\partial \Omega_j$ of a defect $\Omega_j$:

\begin{equation}
 \left. \dfrac{\partial u}{\partial n} \right|_{\partial \Omega_j} = 0 \,, \quad j = 1,...,N
\end{equation}

We would like to consider a particular case when the defects $\Omega_1,...,\Omega_N$ are spherical shells with thin 
water channels connecting a fluid-filled interior cavity to the exterior surrounding fluid. These defects can be 
modelled as multistructures \cite{kmm1999} in the following way,
\begin{equation}
 \Omega_{(N)} = \left\{ a_{(N)} < \sqrt{x^2+y^2+z^2} < b_{(N)} \right\} \setminus \overline{\bigcup_{j=1}^N \Pi_{\varepsilon {\large (N)}}^{(j)}}
\end{equation}
where $a_{(N)}$ and $b_{(N)}$ are given constants and $\Pi_{\varepsilon {\large (N)}}^{(j)}$ is the thin channel. 

\section{\label{sec:level1}Asymptotic approximation: A discrete spring-mass model}

In this section, we derive an asymptotic approximation of the field within thin channels $\Pi_{\varepsilon}^{(j)}$,

\begin{multline}
 \Pi_{\varepsilon}^{(j)} = \left\{ (x,y,z) \, : \, 0 < x < l_j, \right. \\
\,\left. \sqrt{y(t)^2+z(t)^2} < \varepsilon h_j(t), \, (0\leq t \leq 2\pi) \right\}
\end{multline}
where $l_j$ is the length of the $j^{th}$ bridge, $\varepsilon h_j(t)$ the radius of its varying cross-section $D_{\varepsilon}$ 
(parametrized by $t$).
Here, $\varepsilon$ is a small positive non-dimensionnal parameter.\\
To derive the asymptotic expansions, we introduce the scaled variables $\xi = (y/\varepsilon,z/\varepsilon)$. \\


Without loss of generality, and for the sake of simplicity, we drop the
superscript $j$.
In $\Pi_{\varepsilon}$, the time-harmonic wave equation takes the rescaled form

\begin{equation}
 \left\{ \dfrac{1}{\rho}\left( \dfrac{1}{\varepsilon^2} \Delta_{\xi}+\dfrac{\partial^2}{\partial x^2}\right)+\dfrac{\omega^2}{\lambda}\right\}u = 0 \, , 
\end{equation}
with the Neumann boundary conditions

\begin{equation}
 \left. \dfrac{\partial u}{\partial {n}}\right|_{\partial D_{\xi}} = 0
\end{equation}
The field $u$ is approximated in the form

\begin{equation}
u \sim u^{(0)}(x,y,z)+\varepsilon^2 u^{(1)}(x,y,z)
\end{equation}
To leading order, we obtain

\begin{equation}
\left\{ \begin{array}{l l l l}
\Delta_{\xi}u^{(0)} & = & 0 & \quad \text{on} \quad D_\xi\\
\nabla_{\xi}u^{(0)} & = & 0 & \quad \text{on} \quad \partial_{D_\xi}
\end{array} \right.
\end{equation}
Hence, $u^{(0)} = u^{(0)}(x)$ (it is $\xi,\zeta-\text{independent}$). Assuming that $u^{(0)}$ is given, we derive that 
the function $u^{(1)}$ satisfies the following model problem on the scaled cross-section of $\Pi_{\varepsilon}$


\begin{equation}
\left\{ \begin{array}{l l l l}
\Delta_{\xi}u^{(1)} & = & -\dfrac{1}{\rho}\dfrac{\partial^2 u^{(0)}}{\partial x^2}+\dfrac{\omega^2}{\lambda}u^{(1)} & \quad \text{in} \quad D_\xi\\
\nabla_{\xi}u^{(1)}\cdot {\bf n} & = & 0 & \quad \text{on} \quad \partial_{D_\xi}
\end{array} \right.
\end{equation}
The condition of solvability for the problem has the form:

\begin{equation}
 \dfrac{1}{\rho}\dfrac{d^2u^{(0)}}{dx^2}+\dfrac{\omega^2}{\lambda}u^{(0)} = 0 \, , \quad 0<x<l_j
\end{equation}

Hence, we have shown that to the leading order we can approximate the field $u$ within the thin channel $\Pi_{\varepsilon}$ 
by the function $u^{(0)}$ which satisfies the Helmholtz's equation in one-space dimension.\
We now assume that the field is periodic over the cell since it is localized. This shows that the average of the 
eigenfield over the macro-cell vanishes. Indeed, let $\chi_1$ denotes the value of the field in the large body $\Sigma$ of 
the multi-structure $\Omega$ and let $\chi_2$ (which we normalize to 1) denotes the value of the field within the complementary area of the macro-cell $Y \setminus \Omega$ excluding the thin channels. Taking $V=1$ in (\ref{eq3}), 
we deduce that

\begin{equation}
 \omega^2 \int_Y \rho u dx dy dz = -\int_{\partial Y \cup \partial \Omega} \lambda \dfrac{\partial u}{\partial n}dS = 0
\end{equation}
This shows that the average of the field $u$ over $Y$ vanishes, hence by neglecting the volume of the thin channels, 
we obtain

\begin{equation}
 \chi_1 {meas}_{\Sigma} + \chi_2 {meas}_{Y \setminus \Omega} = O(\varepsilon)
\end{equation}
where ${meas}_{\Sigma}$ and ${meas}_{Y \setminus \Omega}$
denote respectively the
areas of $\Sigma$ and $Y \setminus \Omega$.

We now consider two cases. The first one is the study of an array of simple spherical shells with
either one or six thin channels, and the other one is the study
of an array of double spherical shells with one thin channel in each shell.
Since we have $p$ thin channels, we have $p$ separate eigensolutions $V_j,\, (j=1,...,p)$, corresponding 
to the vibrations of thin domains $\Pi_{\varepsilon}^{(j)}$

\begin{equation}
\rho^{-1} V''_j(x)+\lambda^{-1}\omega^2V_j(x) = 0 \, , \quad 0<x<l_j \, , \label{syst_eq_mass1}
\end{equation}

\begin{equation}
V_j(0) = \chi_2 = -\chi_1 \dfrac{meas(\Xi)}{meas(Y \setminus \Omega)} \, ,
\end{equation}

\begin{equation}
\lambda^{-1} I_j V'_j(l_j) = M_j\omega^2V_j(l_j) \, ,  \label{syst_eq_mass3}
\end{equation}
where
\begin{equation}
 I_j=\int_0^{2\pi} \! \! \varepsilon h_j(t) \, dt
\end{equation}
All the channels are connected to $\Xi$, hence, $V_1(l_1) = ... = V_p(l_p) = V$. We note that $V_j(0)$ is equal to a 
non-zero constant.

The solution of the problem $\eqref{syst_eq_mass1}-\eqref{syst_eq_mass3}$ has the form

\begin{equation}
V_j(x) = -\dfrac{\chi_2[\cos((\omega/c)l_j)-1]}{\sin((\omega/c)l_j)}\sin \left( \dfrac{\omega}{c}x \right) + 
\chi_2 \cos \left(\dfrac{\omega}{c}x \right)
\end{equation}
where $c= \sqrt{\lambda/\rho}$ and the frequency $\omega$ is given as the solution of the following equation:

\begin{equation}
\sum_{j=1}^n \left( I_j \cot\left( \dfrac{\omega l_j}{c}\right)\right) = \dfrac{m_jc}{\lambda}\omega \label{aproxeig1}
\end{equation}
where we invoked Newton's second law. Looking at a first low frequency, we deduce an explicit asymptotic approximation

\begin{equation}
\omega \sim \sqrt{\sum_{j=1}^n \left( \dfrac{I_j}{l_j}\right) }\sqrt{\dfrac{\lambda}{M}}
\left(1+\dfrac{meas(\Xi)}{meas(Y \setminus \Omega)} \right) \label{eq20}
\end{equation}

This estimate actually holds for the frequency $\omega_2$ of the upper edge of the phononic band gap. We note that if 
we take $V(0) = 0$ instead of $V(0) = meas(\Xi)/meas(Y \setminus \Omega)$, we estimate the frequency of the 
lower-edge of the phononic band gap.

\subsection{Eigenfrequency estimate in the case of a single spherical shell
with one or six thin channels}
We report in figures (\ref{fig4}) and (\ref{fig6}) finite element computations for a periodic cell of
sidelength $d$ with a simple spherical shell with thin channels. The interior and exterior radii
of the shell are respectively $0.3$m and $0.4$m, the thin channels have the same length
$0.1$m and radii $0.01$m. Therefore, the frequency estimates are (in Hertz):

\begin{figure}[htp]
  \centering
\includegraphics[scale=0.5]{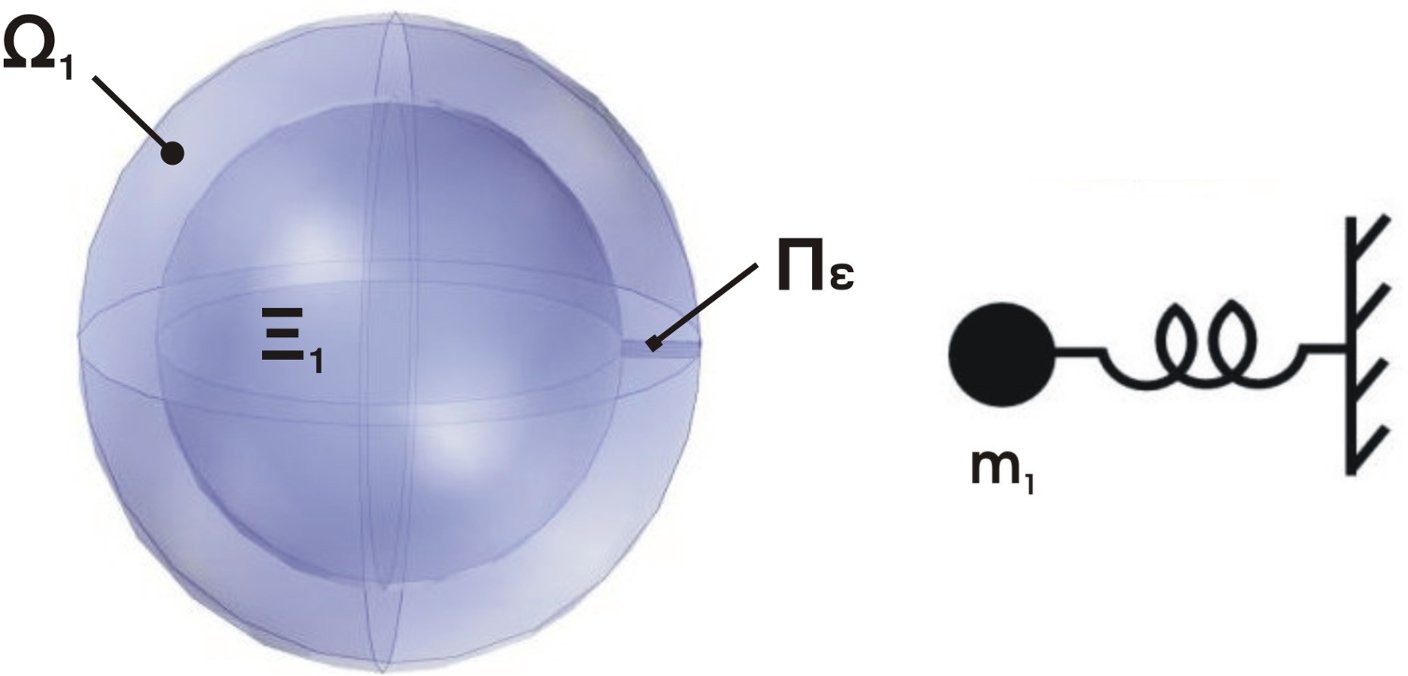} 
\caption{Geometry of the inclusions and the Helmholtz oscillator consisting of one spring connected to a mass at one end 
and fixed at the other end}
\includegraphics[scale=0.17]{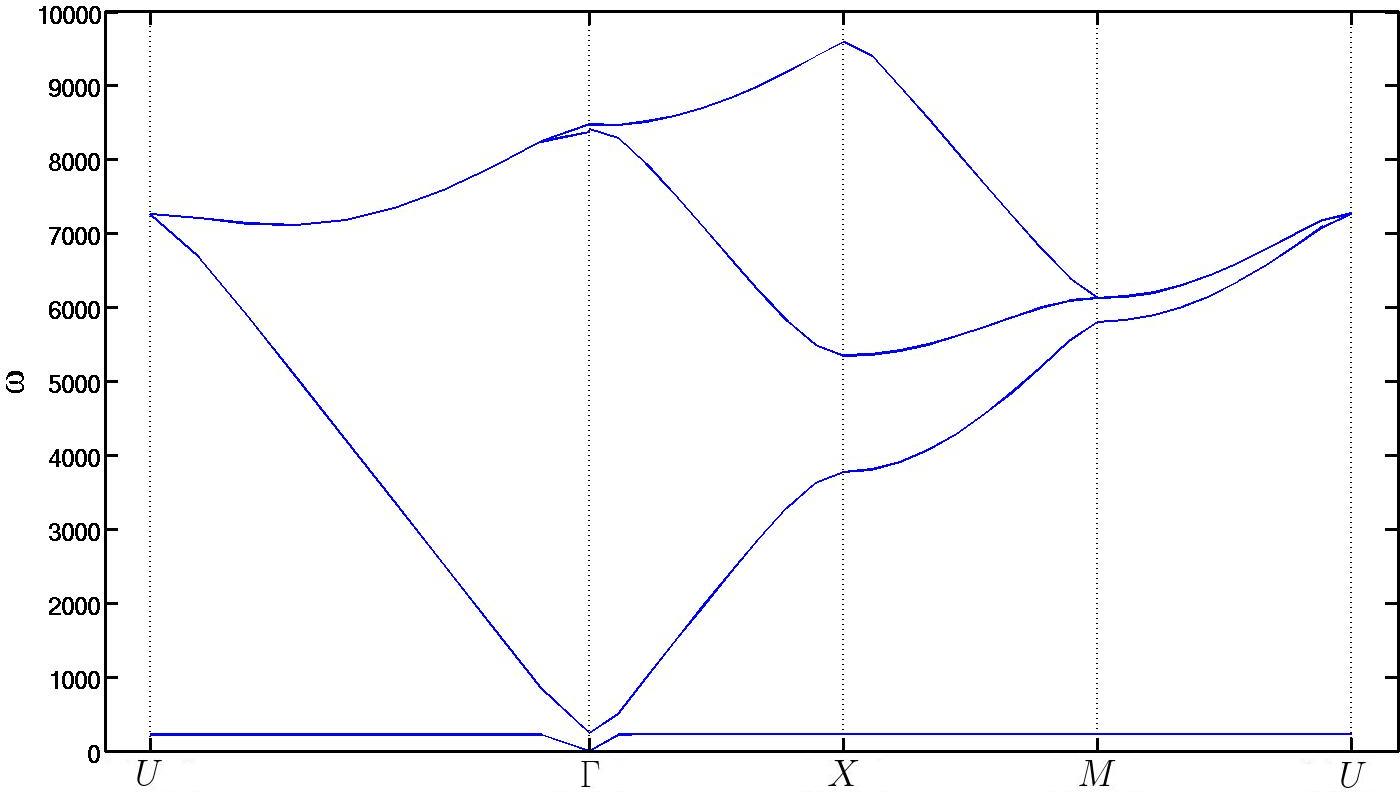} 
\caption{
Dispersion diagram for a periodic array (pitch $d=1$m) of spherical rigid shells (inner radius
$0.3$m and outer radius $0.4$m)
with one thin channel (length $0.1$m and radius $0.01$m) representing the frequency 
$\omega$ (Hz) of pressure waves in water versus the wavenumber $|$\textbf{k}$|$ ($m^{-1}$), projection
of the Bloch vector ${\bf k}$ along the edges of $\Gamma X M U$.
We note the appearance of a frequency stop band for $\omega\in [224,241]$Hz.
}
\label{fig4}
\includegraphics[scale=0.2]{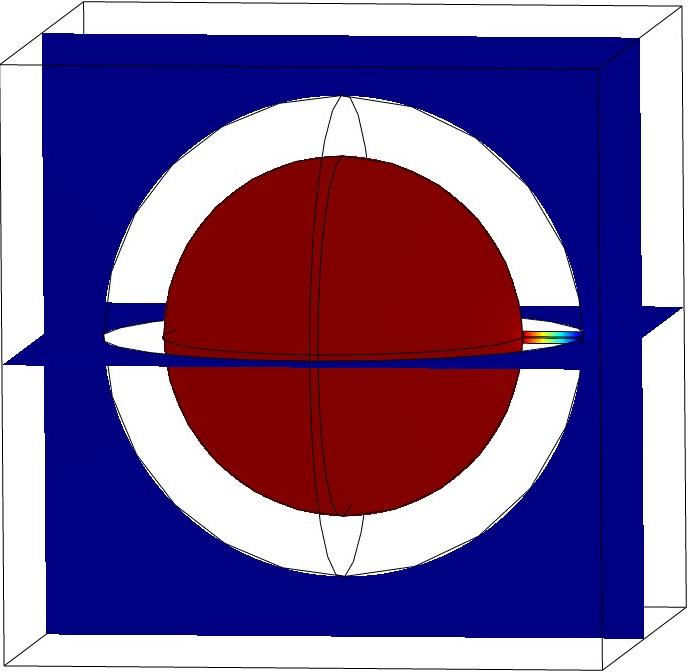}
 \caption{The eigenfunction corresponding to the eigenfrequency
$\omega_1^*=224.6155$Hz for one thin channel. Blue color
corresponds to nearly vanishing amplitude of the eigenmode
$u$, while
red color corresponds to it maximum value.
The pressure field $u$ is constant inside the inner cavity
and outside the shell, but it
varies rapidly inside the thin channel:
it is a localised eigenmode responsible for the
stop band in figure \ref{fig4}, which is
well approximated by
a spring mass model.}
\label{fig5}
\end{figure}

\begin{equation}
 \omega_1 \sim 227.301 \quad , \quad \omega_2 \sim 235.889
\end{equation}
for one thin channel, which are in good agreement with the finite element values

\begin{equation}
 \omega_1^* = 224.6155 \quad , \quad \omega_2^* = 241.3825
\end{equation}
for one thin channel, and

\begin{equation}
 \omega_1 \sim 556.772 \quad , \quad \omega_2 \sim 577.809
\end{equation}
for six thin channels, which are in good agreement with the finite element values

\begin{equation}
 \omega_1^* = 548.3515 \quad , \quad \omega_2^* = 588.8909
\end{equation}
for six thin channels.

\noindent This demonstrates that the discrete model provides
accurate estimates for the lower and upper edges of the
ultra-low frequency stop band. This is therefore
a useful tool which can be used in the design
of acoustic metamaterials.

\begin{figure}[htp]
\includegraphics[scale=0.6]{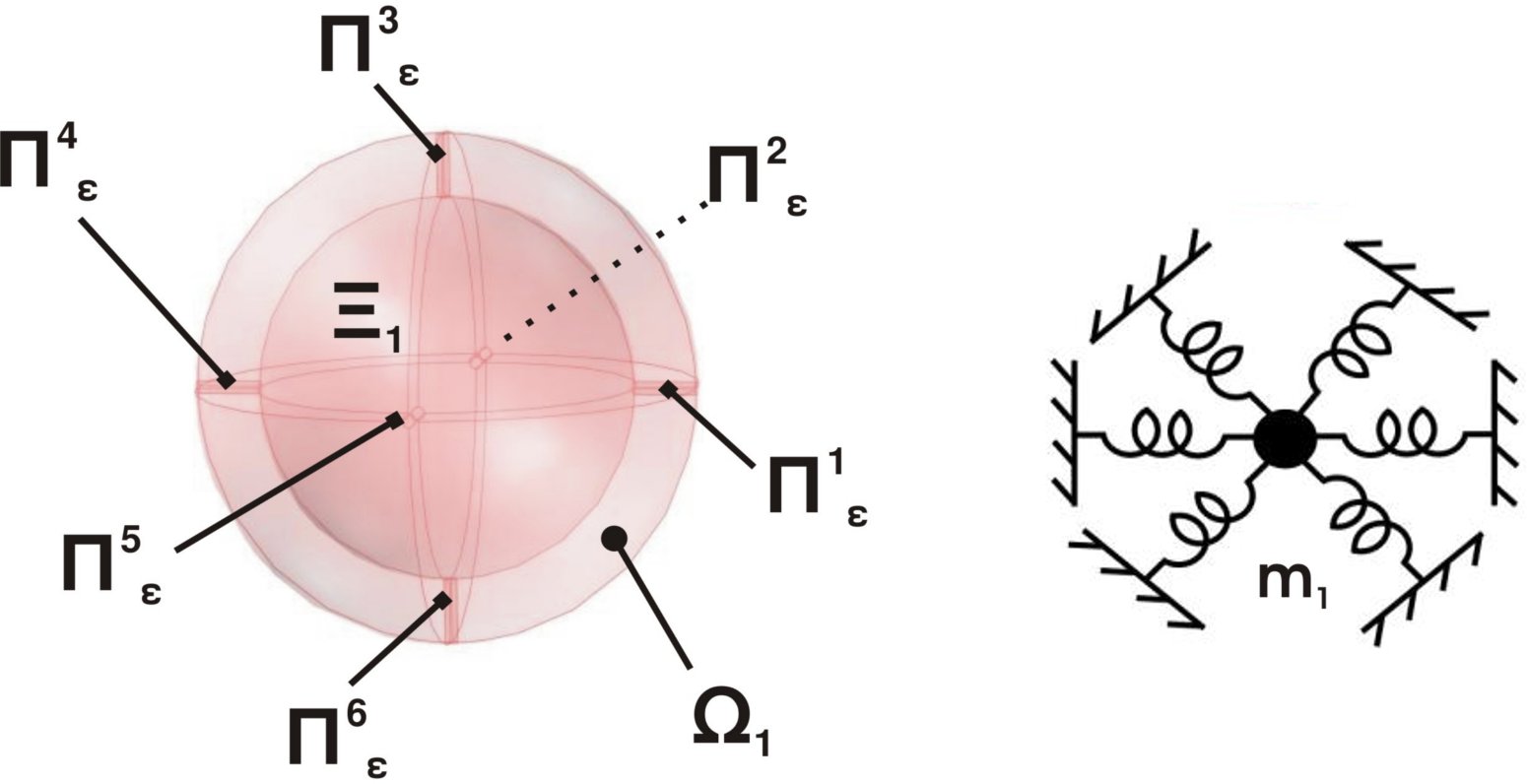}
\caption{Geometry of the inclusions and the Helmholtz oscillator consisting of six springs connected to a mass at one end 
and fixed at the other end}
\includegraphics[scale=0.17]{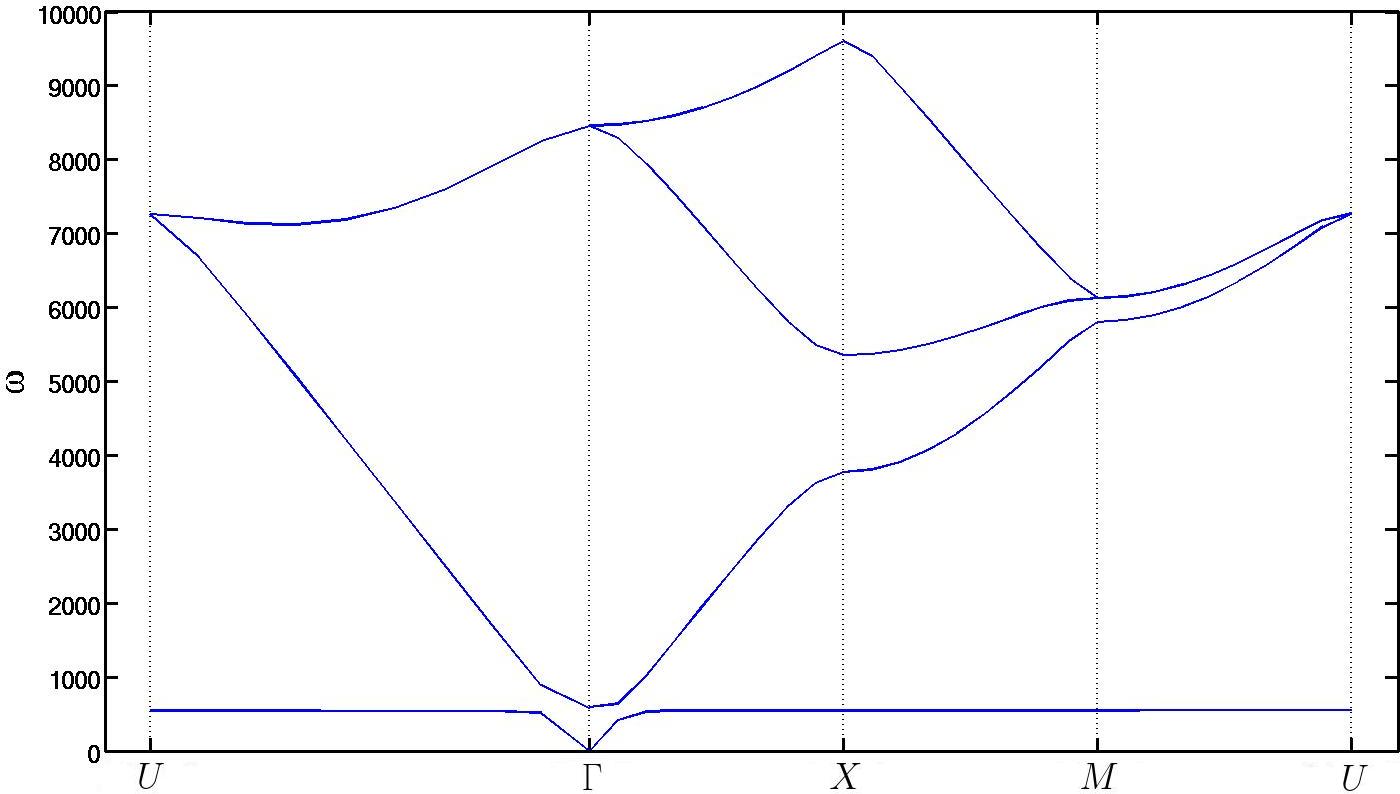}
 \caption{
Dispersion diagram for a periodic array (pitch $d=1$m) of spherical rigid shells (inner radius
$R=0.3$m and outer radius $R=0.4$m)
with six thin channels (length $0.1$m and radius $0.01$m) representing the frequency 
$\omega$ (Hz) of pressure waves in water versus the wavenumber $|$\textbf{k}$|$ ($m^{-1}$), projection
of the Bloch vector ${\bf k}$ along the edges of $\Gamma X M U$.
We note the appearance of a frequency stop band for $\omega\in [548,588]$Hz
which is wider and at higher frequencies than the stop band in figure \ref{fig4}:
the more identical thin channels, the higher the resonant frequency of the localised mode.
} \label{fig6}
\includegraphics[scale=0.2]{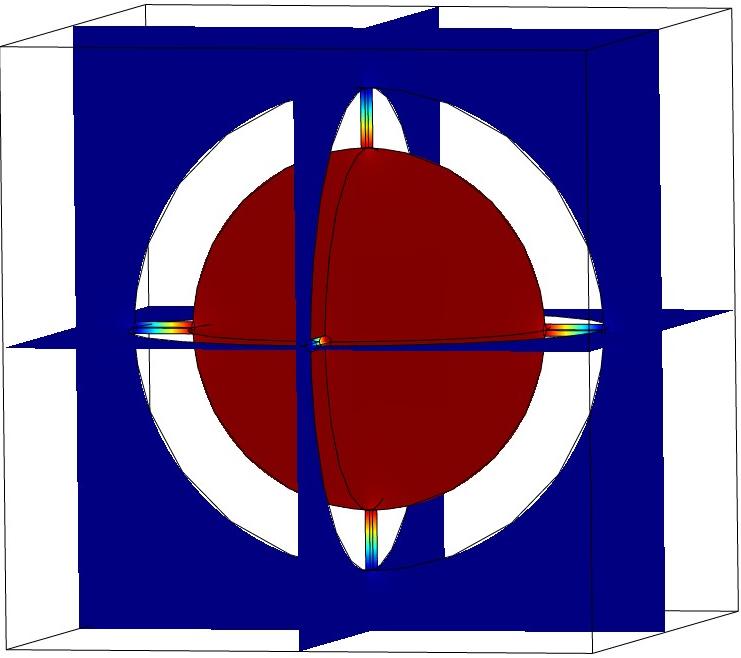}
 \caption{The eigenfunction corresponding to the eigenfrequency $\omega_1^*=548.3515$Hz for six thin channels
responsible for the stop band in figure \ref{fig6}. This frequency is well approximate by the sping mass
discrete model which provides us with the frequency estimate $\omega_1=556.772$Hz.}
\end{figure}

\subsection{Eigenfrequency estimate in the case of a double spherical shell
with one thin channel}
In the numerical example we now have
$\varepsilon^2 h_2^2 = 3.14 .10^{-4}$m$^2$, 
$\varepsilon^2 h_1^2 = 7.85 .10^{-5}$m$^2$, $l_2 = 0.1$m, $l_1 = 0.05$m, and the masses (in kilogram)

\begin{align}
m_1 &= \rho V_1=\dfrac{4000}{3}\pi r_1^3 \nonumber \\
m_2 &= \rho V_2=\dfrac{4000}{3}\pi\left(r_1^3+(b_2^3-a_2^3)\right)+10^3\varepsilon^2 h_1^2l_1 
\end{align}
where $\rho$ is the density of water ($\sim 10^3$kg.m$^{-3}$), $V_1$ and $V_2$ the volumes water occupies in $\Xi_1$ and $\Xi_2$,
$r_1$ is the interior radius for the domain $\Xi_1$ and $a_2$, $b_2$ are respectively the interior 
and exterior radii for the domain $\Xi_2$. In our case, $r_1 = 0.1$m, $a_2 = 0.15$m and $b_2 = 0.2$m.
The formula (\ref{eq20}) gives the following values for the first eigenfrequencies (in Hertz) of the 
multistructures $\Pi_{\varepsilon {\large (1)}} \bigcup \Xi_{(1)}$ et $\Pi_{\varepsilon {\large (2)}} \bigcup \Xi_{(2)}$:

\begin{equation}
 \omega_1 \sim 909.065 \quad ,\quad \omega_2 \sim 542.026
\end{equation}

The corresponding frequencies (in Hertz) associated with the standing waves in the periodic structure were obtained numerically, and 
they are

\begin{equation}
 \omega_1^* = 979.0268 \quad , \quad \omega_2^* = 470.286 
\end{equation}

\begin{figure}[htp]
  \centering
\includegraphics[scale=0.3]{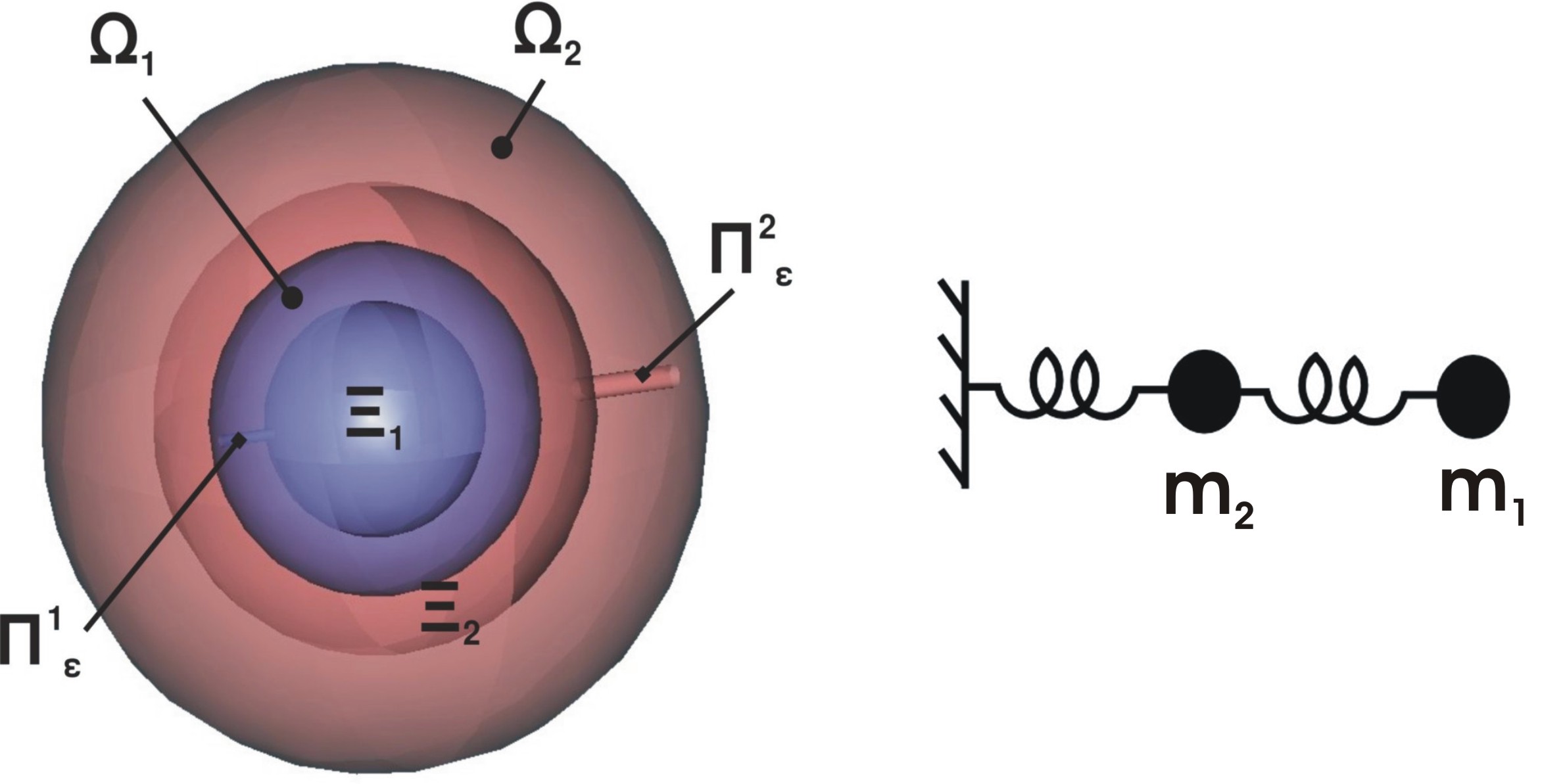}
\caption{Geometry of the inclusions and the Helmholtz oscillator consisting of two masses connected by a spring and one 
connected to a fixed domain.}
\includegraphics[scale=0.18]{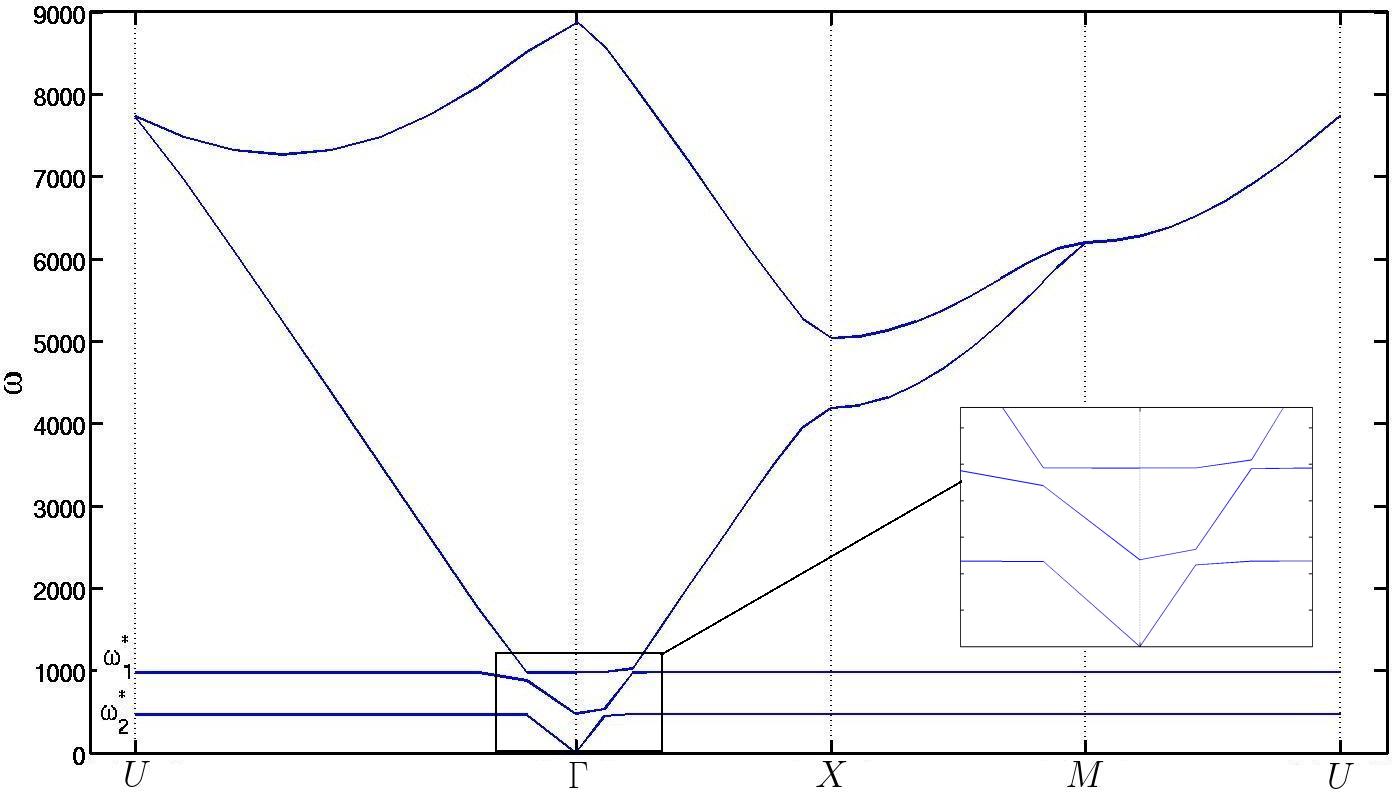} 
  \caption{
Dispersion diagram for a periodic array (pitch $d=1$m) of double spherical rigid shells (radius
of spheres from inner to outer are $0.1$m, $0.15$m, $0.3$m and $0.4$m)
with one thin channel in each shell (respectively of lengths $0.05$m and $0.1$m and radii $0.005$m and $0.01$m)
representing the frequency 
$\omega$ (Hz) of pressure waves in water versus the wavenumber $|$\textbf{k}$|$ ($m^{-1}$), projection
of the Bloch vector ${\bf k}$ along the edges of $\Gamma X M U$.
We note the appearance of two frequency stop bands for $\omega\in [470.2860,476.5000]$Hz
and $\omega\in[979.0268,979.3559]$Hz. 
}
\includegraphics[scale=0.2]{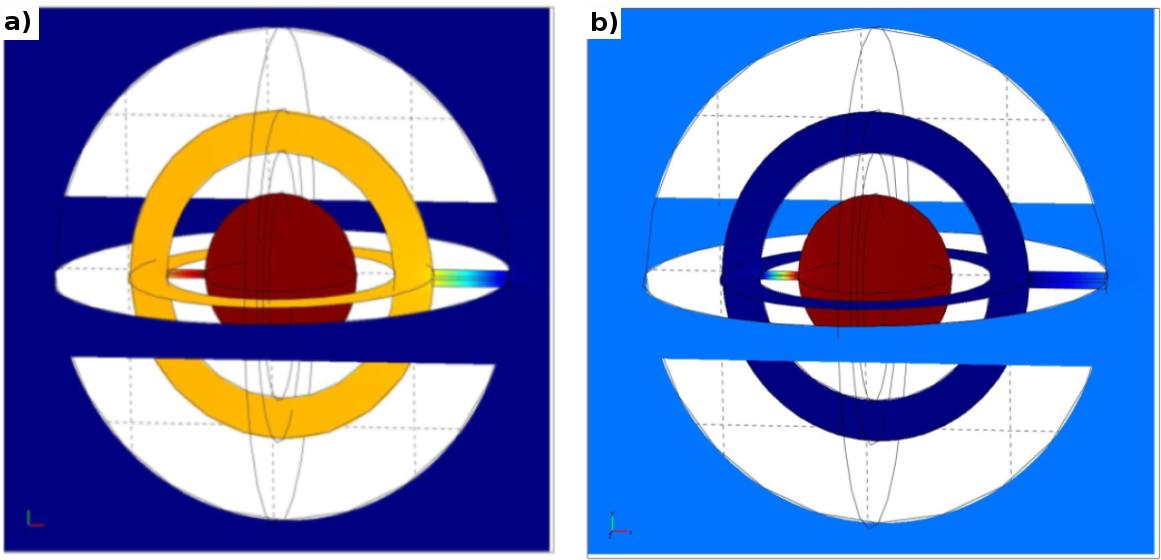}
 \caption{a) The eigenfunction corresponding to the eigenfrequency $\omega_1^*=979.0268$Hz.
b) The eigenfunction corresponding to the eigenfrequency $\omega_2^*=470.286$Hz.
In (a) both canals vibrates, while in (b) only the inner canal does.
Both frequencies are reasonably
well approximated by the spring mass model
which gives $\omega_1=909.065$Hz and $\omega_2=542.026$Hz.
}
\end{figure}

Formula (\ref{eq20}) gives a good estimate for the eigenfrequency $\omega_1^*$, but we observe a discrepancy in the approximation of $\omega_2^*$. The estimate for the eigenfrequency $\omega_2^*$ can be improved if the domain 
$\Pi_{\varepsilon {\large (2)}} \bigcup \Xi_{(2)}$ is replaced by the domain 
$\Pi_{\varepsilon {\large (2)}} \bigcup \Omega_{(2)} \bigcup \Pi_{\varepsilon {\large (1)}} \bigcup \Omega_{(1)}$.
In this case, the eigenfrequency $\omega_2$ is approximated by the first positive eigenvalue of the problem

\begin{equation}
\rho^{-1} V''_1(x)+\lambda^{-1}\omega^2V_1(x) = 0 \, , \quad 0<x<l_1
\end{equation}

\begin{equation}
V_1(0) = 0 \, ,
\end{equation}

\begin{equation}
\lambda^{-1}I_1 V'_1(l_1) -\lambda^{-1}I_2 V'_2(0) 
= m_1\omega^2V_1(l_1) \, ,
\end{equation}

\begin{equation}
\rho^{-1} V''_2(x)+\lambda^{-1}\omega^2V_2(x) = 0 \, , \quad 0<x<l_2
\end{equation}

\begin{equation}
\lambda^{-1}I_2 V'_2(l_2) = m_2\omega^2V_2(l_2) \, ,
\end{equation}

\begin{equation}
V_2(0) = V_1(l_1) \, ,
\end{equation}
where $V_1(x)$, $V_2(x)$ are the eigenfunctions defined on $(0,l_1)$ and $(0,l_2)$, respectively, and the masses 
$m_1$, $m_2$ are defined by (in kilogram)

\begin{align}
m_1 &= \dfrac{4000}{3}\pi r_1^3 \nonumber \\
m_2 &= \dfrac{4000}{3}\pi\left(b_2^3-a_2^3\right) 
\end{align}
Taking into account that $\omega_2 = O(\varepsilon)$, we deduce that it can be approximated as the first positive 
solution of the following algebraic equation:

\begin{multline}
m_1m_2l_1l_2\omega^4- \lambda \omega^2 \left( l_2I_1m_1+I_1l_2m_2+I_2m_1l_1 \right) \\
+\lambda^2I_1I_2 = 0 
\end{multline}
so that $\omega_2 \sim 520.121$Hz, which provides a more accurate approximation
of $\omega_2^*=470.286$H.

\section{Conclusion}
\begin{figure}[htp]
  \centering
\includegraphics[scale=0.3]{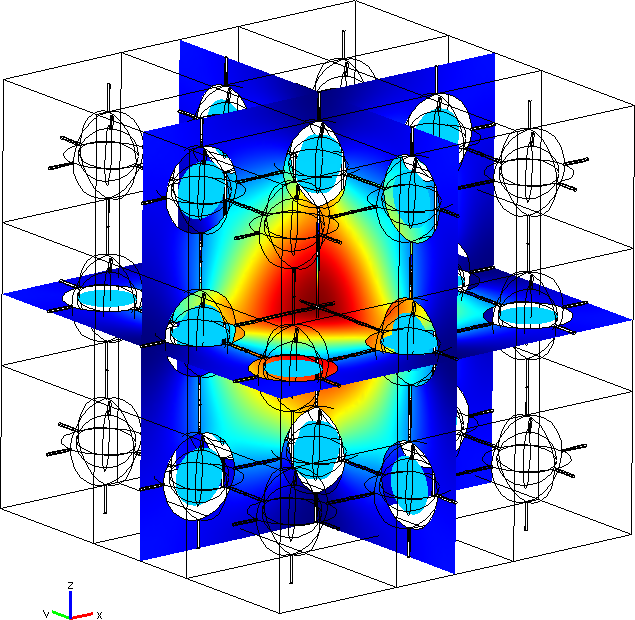}
 \caption{The eigenfunction corresponding to the eigenfrequency $\omega_1^*=224.6155$Hz
for a macrocell of 26 resonators as in figures \ref{fig4} and \ref{fig5} with a defect
(fluid instead of resonator) in the middle. The localised mode sits within
the ultra-low frequency stop band of figure \ref{fig4}.}
\label{fig12}
\end{figure}

In this paper, we have seen that it is possible to sculpt the Bloch spectrum of three-dimensional phononic
crystals ad libitum simply by digging some holes and adding cavities in rigid spheres periodically arranged along a cubic lattice. One of the
main achievements of the numerical study is the appearance of ultra-low frequency stop bands at frequencies predicted quantitalively by an asymptotic model. We also conducted sozme basic shape optimization (by varying the size, diameter and number of channels in a
rigid sphere of constant radius) in order to
enhance the control of the location and the number of low frequency stop bands, thanks to our asymptotic estimates.
Importantly, a cubic array of rigid spheres does not support any complete stop band, even in the
densely packed configuration. Our findings are thus twofold: Multistructures open not only an original route toward ultra-low frequency stop bands (associated with very flat dispersion curves i.e. localized eigenmodes) but also offer the first paradigm of a complete stop band for three-dimensional pressure waves in a fluid. Finally, we illustrate in figure \ref{fig12} a possible application of the ultra-low frequency
stop band in order to localize a mode of a wavelength much larger than the pitch of the array of resonators. Similarly, one could envisage to reflect, detour, or focus, pressure waves using the low frequency stop band within which effective parameters are expected to take negative values as it is now well-established for the two-dimensional counterpart of such kind of acoustic metamaterials.

\end{document}